# Language Models Reproduce Human Reductionist Bias and Decision Inconsistency in Neurodevelopmental Disorders Assessment

**Maciej Wodziński[1], Joanna Wodzińska[1], Kacper Dudzic[2,3,4], Marcin Moskalewicz[1,2,3]**

[1] Maria Curie-Skłodowska University, Plac Marii Curie-Skłodowskiej 4, 20-031 Lublin, Poland
[2] IDEAS Research Institute, Krakowskie Przedmieście 13, 00-071 Warsaw, Poland
[3] Poznan University of Medical Sciences, Collegium Maius, Fredry 10, 61-701 Poznan, Poland
[4] AMU Center for Artificial Intelligence, Poznan, Poland

**Abstract**

Large language models (LLMs) are increasingly supporting complex mental-health decisions, which depend not only on factual evidence but also value-laden interpretations. We introduce a mixed-methods human–LLM auditing framework examining decision consistency, susceptibility to cognitive heuristics, declarative intellectual humility, and the concepts operationalized in support-allocation judgments of neurodevelopmental disorders. Comparing 35 humans (18 physicians and 17 psychologists) with seven LLMs, we show that in both groups, ratings of patients' functional level were not significantly associated with support-eligibility decisions, indicating an inconsistency between descriptive assessments and final evaluative judgments. Specifically, we find that neither group showed significant susceptibility to experimental manipulations targeting anchoring and representativeness heuristics. LLMs reported higher intellectual humility than experts ($U = 241$, $p < .001$, $r = .62$; LLMs: $M = 41.43$, $SD = 1.99$; experts: $M = 29.03$, $SD = 8.05$), but it was unrelated to decision consistency or functional assessment. While LLMs and physicians granted support less frequently than psychologists ($U = 180.50$, $p = .003$, $r = .34$), they also interpreted a concept of "basic life needs" differently, primarily as biological survival and self-care, and not communicative and social needs. These findings suggest that despite expressing high levels of intellectual humility, LLMs reproduce a reductionist interpretive framework and knowledge embedded in medical decision-making. More broadly, we argue that evaluating AI in high-stakes contexts requires not only measuring accuracy, agreement, or resistance to cognitive bias, but also critical examination of the concepts of neurodiversity that AI systems operationalize.

## Introduction

The increasingly widespread use of large language models (LLMs) as tools to support medical decision-making means that testing them across various contexts is particularly important (Cross, Choma, and Onofrey 2024; Ke et al. 2024). This is not only due to the architecture and design of the models, but above all to the quality of the results they produce and their tendency to be subject to heuristics and biases, including those characteristic of humans (Mahajan et al. 2025). Our experiment compares the decisions made by selected LLMs with those made by mental health experts (psychologists and psychiatrists). The questions posed to both groups of experts and to the LLMs were based on those asked of experts during Polish Disability Assessment Boards (DAB) and were also designed to test whether the participants were subject to representativeness and anchoring heuristics (Richie and Josephson 2018; Sætrevik et al. 2024). In Poland, disability is formally assessed by district or municipal Disability Assessment Boards (DAB), with regional boards acting as appellate bodies. These multidisciplinary institutions include physicians, psychologists, educators, vocational counselors, and social workers because their task is not merely to confirm a diagnosis, but to translate medical, psychological, and functional evidence into an administrative judgment concerning the person's disability and support needs. DAB's decisions are particularly consequential for people with neurodevelopmental conditions, as a clinical diagnosis alone does not automatically establish disability, while autism- or ADHD-related difficulties may be contextdependent, uneven across domains, and insufficiently visible during a brief assessment. A DAB meeting is an ideal setting for studying systemic cognitive errors, which are likely to occur. Systemic cognitive errors are individual biases that recur across decisionmakers and may be reinforced by institutional procedures, professional norms, and shared interpretive frameworks (O'Sullivan and Schofield 2018; Tversky and Kahneman 1974), with mental health experts being particularly susceptible to them (Shanteau 1992; Crumlish and Kelly 2009). This is because the assessment is usually issued within a short timeframe (sometimes in under 10 minutes), experts receive information from various sources (their own observations, medical records, from the family or the patient), the judged cases can be highly complex, and the observational examination takes place in artificial, almost laboratory-like conditions that do not reflect the realities faced by the individual in their everyday life. The resulting certificate provides access to a range of benefits, services, accommodations, and rehabilitation or employment-related entitlements; an inadequate assessment may therefore restrict access to support essential for everyday functioning and social participation. Previous studies suggest that these decisions may be shaped by limited condition-specific knowledge (Wodziński, Rządeczka, and Moskalewicz 2023), stereotypical or reductionist understandings of functioning (Wodziński and Moskalewicz 2023), and the insufficient recognition of testimony provided by neurodivergent people and their families (Krawczyk et al. 2025). One of the most difficult questions that the DAB must answer during a hearing – and, in the view of the disability community, the most controversial – is: "Whether the individual requires constant or long-term care or assistance from another person due to a substantially limited capacity for independent living?". We further refer to this as 'Point 7', based on the reference number specifying the entitlements granted to a particular person on their disability certificate. This term is already widely used within the disability community to describe the problem of being granted the appropriate benefits. The implementing regulations for this decree clarify that "Limited capacity for independent living means an impairment of bodily function to such an extent that it prevents the person from meeting their basic life needs without the assistance of others (...)". A question formulated in this way requires interpretation, based on the expert's knowledge and experience, regarding what constitutes 'the ability to live independently', what exactly 'basic living needs' are, and when they can be considered to have been met. Recognising that the person being assessed requires such assistance means granting them and their family significant resources, often essential for day-to-day functioning, for example, where one parent has to give up paid work to act as a carer. We have chosen these committees as an example because they bring into sharp focus the problems associated with assessing the condition of patients with neurodevelopmental disorders. Furthermore, although the example of the DAB committees is drawn from the Polish cultural context, it illustrates a broader phenomenon of difficulties and cognitive biases to which health experts are exposed when making decisions under conditions of high cognitive uncertainty, with limited time and a wide range of possible interpretations of data obtained from many different sources (Ly, Shekelle, and Song 2023; Reale et al. 2023). The field of psychiatry is particularly susceptible to cognitive biases — in both humans (Abi-Dargham et al. 2023) and LLMs (Bouguettaya, Stuart, and Aboujaoude 2025) — due to the lack of biological markers associated with the diagnoses in question and the fact that medical decisions are based primarily on descriptions of behaviour and the presence of symptoms codified in diagnostic manuals, and consequently involve a high degree of subjectivity in the decision-making process. Moreover, DAB judgments may serve as a testbed for the problem of balancing different contexts of human functioning, which is addressed in international assessment standards. The prime example of these standards, the WHO International Classification of Functioning, Disability and Health (ICF), mandates a biopsychosocial assessment against a purely medical model, balancing body functions, activities, and participation in life situations (WHO 2001; Lutte et al. 2024). We wanted to investigate whether the language models' decisions would be more consistent than human professionals', how they understand the basic concepts related to assessing the level of functioning of patients with neurodevelopmental disorders, and, consequently, what criteria they take into account during the decision-making process.

## Methods

### Research protocol

The research protocol involved two groups of participants: human experts (N=35) and 7 selected LLMs carrying out three tasks. Experts Experts, including medical doctors (N=18) and psychologists (N=17), were recruited via an advertisement posted on social media and through the authors' own networks. Recruitment was challenging because disability assessors are appointed per session, and no accessible sampling frame exists. Recruitment was challenging because disability assessors are appointed to individual sessions rather than permanently employed by local Disability Assessment Boards, leaving no accessible sampling frame. Despite invitations distributed through the boards and financial compensation, only 35 eligible experts volunteered during the recruitment period. The group included 11 males and 25 females with 1 to 43 years of experience (average of 13.3 years) and between 0 and 24 years (average 4.2 years) of direct experience in DAB assessment. They received a fee of approximately 130 EUR for taking part in the study. Tasks 1 and 2 were completed via online forms, whilst Task 3 was carried out using the Zoom platform. Models We evaluated 7 models from 5 leading providers: GPT-5.5 from OpenAI, Claude Opus 4.7 and Claude Sonnet 4.6 from Anthropic, Gemini 2.5 Pro and Gemini 3.1 Pro from Google, as well as Llama 4 Maverick from Meta, and Qwen 3.6-27B from Alibaba Cloud. Our model selection criteria encompassed public interest and general performance as well as availability on the OpenRouter API interface 1. In total, 462 API calls through OpenRouter were performed: (20 cases + 2 qualitative questions) x 3 iterations x 7 models) with the default temperature. The calls contained just the task instructions inserted as the user prompt. No in-context learning paradigm was employed in the inference protocol. The paradigm of conducting several iterations with the default model temperature instead of a single one with a temperature of 0 was deliberate, as we intended to probe the opinion consistency of the models - higher temperature should not be enough to flip the answers given by the models, assuming internal consistency of their opinions, preferences, and biases (Renze 2024; Schroeder and Wood-Doughty 2025). The full protocol for experts and models, including content of all tasks, along with instructions and prompts for the models, can be found in the supplementary materials.

### Tasks

**Task 1: Case studies** The first task involved reading case studies containing fictional descriptions of how people with various types of neurodevelopmental disorders, such as autism, ADHD, or dyslexia, function. Deception was used in the experimental procedure so as not to lead the evaluators to the research objectives of the project. Immediately after reading each case description, and without awareness that it is fictional, the assessor's task was to answer two questions: 1) "I assess the person's level of functioning as:" (possible answers: 1 – significantly impaired, 2 – impaired, 3 – average, 4 – good, 5 – very good) and 2) "Does the person described require constant or long-term care or assistance from another person due to a significantly limited ability to live independently?" (Yes/No). The second question corresponds to "point 7" in the Polish disability certification system. A "Yes" response indicates that the person meets the condition specified in point 7, which may entitle the person to certain forms of disability certification benefits under the Polish system. Two experienced assessors—a clinical psychologist and a special needs teacher—developed vignettes with multimodal manipulations testing anchoring and representativeness. Anchoring occurs when early information disproportionately influences judgment (Kahneman and Tversky 1973), whereas representativeness involves judging cases by their similarity to familiar stereotypes (Kahneman and Tversky 1972); thus, information order or photographs may affect an assessment even when neither should be relevant to the substantive decision. Experts were randomly assigned in an approximately equal ratio to two counterbalanced sequences (n=17 and n=18), each comprising ten cases across two conditions, while each model assessed both versions of all cases (20 total). In the anchoring control condition, positive and negative information was evenly distributed (Fig. 1).

**Boy, aged 10**

The child's functioning reveals diverse developmental difficulties in the cognitive domain. The boy has a speech development disorder. His articulation is incorrect both for individual speech sounds and for whole words. He changes the order of sounds and syllables and adds or "loses" parts of word structure. He requires speech and language assessment and therapy. The symptom that concerned his parents was very severe difficulty in learning to read. Assessment of phonological functions revealed major disturbances.

Phonological-awareness tasks were entirely impossible for him to perform. Phonemic analysis and synthesis pose an enormous problem. Difficulties are evident in every type of task: identifying the first or last sound and combining even two sounds into a syllable. Consequently, the boy will be able to master reading only at a very basic level.

The boy has difficulty visually searching a larger perceptual field and does not do so in a planned and systematic manner. This indicates impaired selectivity of visual attention. Planning processes involved in visual analysis and synthesis are disturbed. This is an executive-function deficit rather than a perceptual deficit.

The boy displays impaired dynamic praxis, that is, difficulty performing rapid movements in a specified sequence. The precision of his motor activities is likewise below the level expected for his age. Graphomotor activities are performed with excessive pressure on the pen. Visuomotor memory for letter-based material is substantially reduced. The dissociation between memory for letters and memory for geometric or pictorial material suggests a primary deficit in language functions.

Visual perception functions normally, both in its analytic-synthetic aspect and in simultaneous gnosis: the boy correctly perceives figure and ground and performs without error tasks involving recognition of overlapping, incomplete, and poorly visible figures against a distracting background. His visual memory for geometric material is very good.

With respect to other auditory functions, the boy hears differences between pairs of words containing similar sounds, which means that his phonemic hearing functions normally. He also copes well with analysis and synthesis at the syllabic level. The positive effects observed from rehabilitation may suggest that he responds well to therapy.

At sentence and text level, his use of linguistic, morphological, and syntactic rules is age-appropriate. Minor errors occur occasionally, but are not significant when his overall development is considered. Auditory memory for non-verbal material is age-appropriate. He correctly names structures and accurately identifies the number of sounds and the pauses between them. The boy is very persistent. He works calmly and, within the limits of his abilities, for long periods (even two hours with a short break). This is one of his strengths and an important positive characteristic useful in school learning.

**Figure 1:** The anchoring heuristic, control group. An example of the even distribution of negative descriptions in a case study (here marked in red).

In the experimental group, negative information (i.e., descriptions of functional difficulties and deficits) was concentrated at the beginning of the case description (Fig. 2).

**Boy, aged 10**

The child's functioning reveals diverse developmental difficulties in the cognitive domain. By contrast, his emotional and social development is proceeding normally.

The boy has a speech development disorder. His articulation is incorrect both for individual speech sounds and for whole words. He changes the order of sounds and syllables and adds or "loses" parts of word structure. He requires speech and language assessment and therapy.

On the other hand, at sentence and text level, his use of linguistic, morphological, and syntactic rules is age appropriate. Minor errors occur occasionally, but are not significant when his overall development is considered.

The symptom that concerned his parents was very severe difficulty in learning to read. Assessment of phonological functions revealed major disturbances. Phonological-awareness tasks were entirely impossible for him to perform. Phonemic analysis and synthesis pose an enormous problem. Difficulties are evident in every type of task: identifying the first or last sound and combining even two sounds into a syllable. Consequently, the boy will be able to master reading only at a very basic level. With respect to other auditory functions, the boy hears differences between pairs of words containing similar sounds, which means that his phonemic hearing functions normally. He also copes well with analysis and synthesis at the syllabic level. The positive effects of rehabilitation may suggest that he responds well to therapy.

Auditory memory for non-verbal material is age-appropriate. He correctly names structures and accurately identifies the number of sounds and the pauses between them.

The boy has difficulty visually searching a larger perceptual field and does not do so in a planned and systematic manner. This indicates impaired selectivity of visual attention. Planning processes involved in visual analysis and synthesis are disturbed. This is an executive-function deficit rather than a perceptual deficit.

Visual perception functions normally, both in its analytic-synthetic aspect and in simultaneous gnosis: the boy correctly perceives figure and ground and performs without error tasks involving recognition of overlapping, incomplete, and poorly visible figures against a distracting background.

His visual memory for geometric material is very good.

The boy displays impaired dynamic praxis, that is, difficulty performing rapid movements in a specified sequence. The precision of his motor activities is likewise below the level expected for his age. Graphomotor activities are performed with excessive pressure on the pen.

The boy is very persistent. He works calmly and, within the limits of his abilities, for long periods (even two hours with a short break). This is one of his strengths and an important positive characteristic useful in school learning.

**Figure 2:** The anchoring heuristic, experimental group. An example of the focused distribution of negative descriptions in a case study (here marked in red).

A rater susceptible to this heuristic could anchor their judgment on either the negative information presented initially or the positive information encountered later, thereby underestimating or overestimating the individual's level of functioning (question 1). This design therefore allowed us to examine whether the ordering of otherwise identical information systematically shifted the relative weight assigned to the person's strengths and difficulties. Such anchoring would consequently result in support being granted more or less frequently (question 2), despite the overall informational content remaining unchanged.

In the case of the representativeness heuristic, the manipulation involved the type of photograph attached to the same case description. In the control group, the photograph depicted a person in a neutral or positive emotional state, whilst in the experimental group, the photograph depicted a person displaying a negative emotional state (Fig. 3).

**Girl MN, born 29 August 2013**

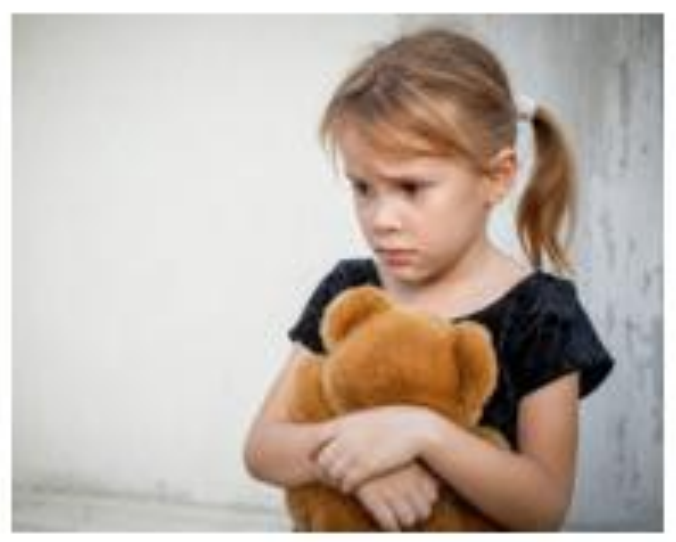

The patient displays atypical behaviour in relationships. She can often be observed isolating herself and appears not to notice the adult. Frequent, firm interventions are needed to maintain the child's attention. She initiates minimal eye contact. The patient's emotional response is very rarely appropriate to the situation. It is difficult to change her mood. She avoids eye contact and often stares into space.

**Girl MN, born 29 August 2013**

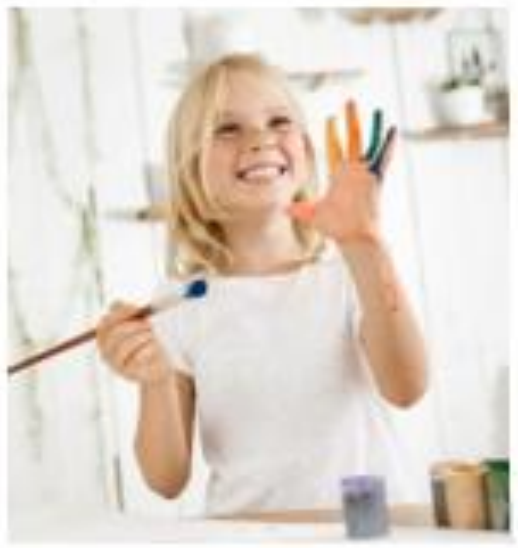

The patient displays atypical behaviour in relationships. She can often be observed isolating herself and appears not to notice the adult. Frequent, firm interventions are needed to maintain the child's attention. She initiates minimal eye contact. The patient's emotional response is very rarely appropriate to the situation. It is difficult to change her mood. She avoids eye contact and often stares into space.

**Figure 3:** The representativeness heuristic. Comparison of different photographs used with the same case description fragments. The experimental group (top) vs. the control group (bottom).

This manipulation tested whether negative affect would be treated as a visual proxy for disability severity, despite providing no additional relevant information about the person's actual functioning or support needs. It also enabled us to examine whether human experts and multimodal LLMs would integrate textual and visual evidence consistently or allow a salient but diagnostically irrelevant image to override the standardized case description. If participants were subject to the heuristic, experimental cases should be rated lower and support granted more frequently than in control cases.

**Task 2: Specific Intellectual Humility Scale** Specific Intellectual Humility Scale (SIHS) questionnaire was administered to assess a declarative level of domain-specific intellectual humility (IH), ie., one's willingness to recognize that his/her belief may be mistaken and to reconsider it in light of new evidence or arguments (Bąk, Wójtowicz, and Kutnik 2022; Bak and Kutnik 2021). Participants rated their answers to 9 questions on a scale of 1 to 5 (1 = I disagree with the statement, 5 = I fully agree with the statement). The total score from these answers indicates the participant's reported level of IH. A low IH score may indicate excessive rigidity and overconfidence in one's own beliefs, leaving the decision-maker vulnerable to errors, including cognitive biases. However, an excessively high level of IH – that is, the view that every belief is equally important – can lead to decision-making paralysis and poor-quality, and sometimes even random, explanations for one's own beliefs. In the case of expert assessments, a tendency towards critical analysis of the available data should go hand in hand with greater consistency in the decisions made (Bowes, Ringwood, and Tasimi 2024). In the case of language models, human selfreport scales likely measure sycophancy and safety training to project modesty, and not true epistemic caution, but nevertheless can be used to measure alignment between declarative humility and behavioral performance (Plisiecki et al. 2026).

**Task 3: Open-ended questions** The final task was to answer open-ended questions. In the case of the experts, this took place during an interview with the researcher, whilst the models were provided with the text of the questions and instructions regarding the format of their responses in the questionnaires sent to them. The interviews lasted between 30 and 60 minutes. The responses were coded independently by two researchers according to the thematic analysis method (Braun and Clarke 2006, 2021), and the final interpretation was established through consensus discussion. The questions most relevant to the issue under investigation were as follows: "What exactly is assessed when determining a person's degree of disability?" "Explain what the phrase 'ability to live independently' means in the sentence 'requires constant or long-term care or assistance from another person due to a significantly limited ability to live independently'", and "Explain what the phrase 'basic living needs' means". The full list of questions can be found in the research protocol in the supplementary materials. This task aimed to use a self-explanation technique (Madsen, Chandar, and Reddy 2024; Marasovic et al. 2022) to gain a more detailed understanding of how the most important concepts are conceptualised and operationalised, and to identify the factors that were taken into account when decisions were made. The responses to the open-ended questions were analysed using thematic analysis to identify the core, recurring categories related to the key concepts of the research area. Community involvement All authors of this paper are active neurodivergent researchers. Participatory research increasingly recognizes that neurodivergent people contribute distinctive epistemic perspectives grounded in lived experience (Cunff et al. 2023; Pickard et al. 2022). Such situated knowledge can reveal interpretive assumptions overlooked by exclusively neurotypical research teams and support greater epistemic justice by treating neurodivergent people as knowledge producers rather than merely research subjects.

## Results

Due to unequal group sizes and the non-normal distribution of most variables, both nonparametric and parametric analyses were conducted, including Mann–Whitney U tests, Kruskal–Wallis H tests, $\chi2$ tests, Spearman's ρ correlations, and independent-samples t tests. For multiple comparisons (factors: source and group), a significance level of $\alpha \leq .025$ was adopted, whereas $\alpha \leq .05$ was used for all other analyses.

### Quantitative analyses

**Resistance to heuristics** Much like the experts, the language models were resistant to the experimental manipulation with respect to their decision-making. Across both heuristics, the experimental manipulation did not produce significant effects, either among expert groups or within the language models, regardless of source type, the number of assigned "point 7" granting disability certification benefits (Fig. 4), or the assessment of the level of functioning.

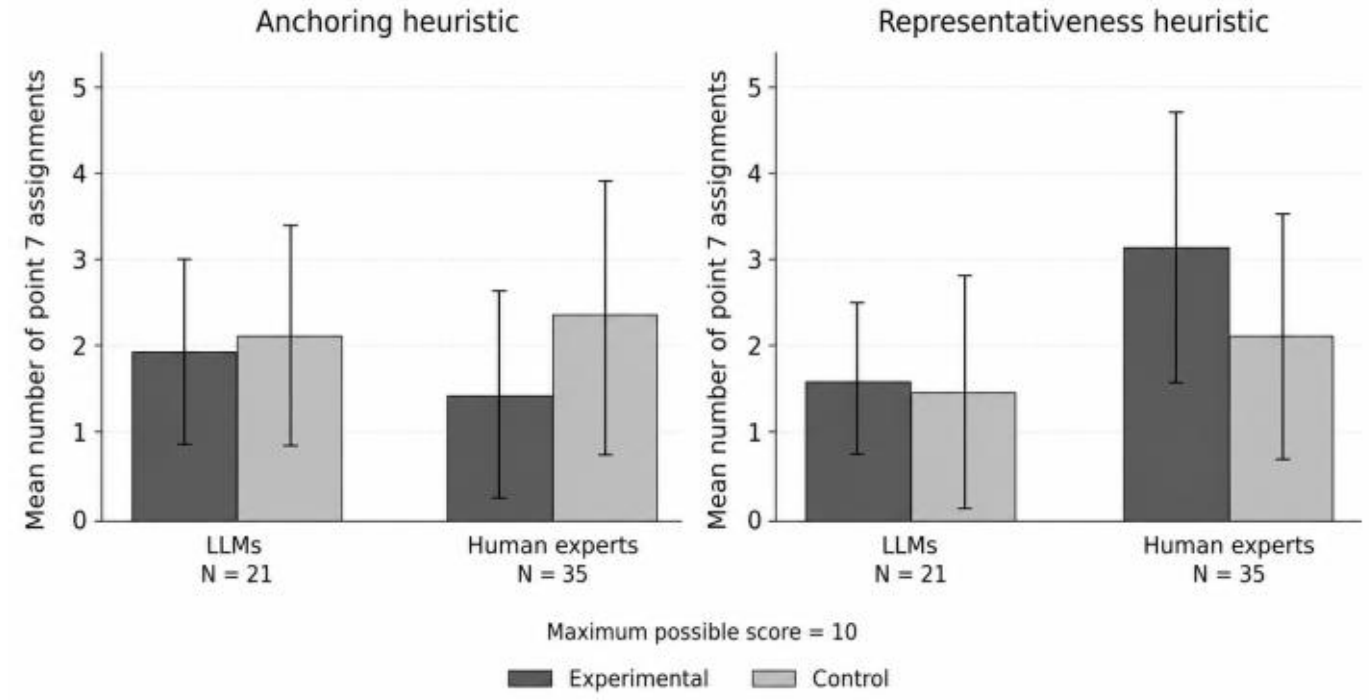

**Figure 4:** Assignments of Disability Certification Benefits ("Point 7") Across Experimental Conditions for Experts and Language Models.

**Lack of consistency** Regardless of the arrangement of information presentation or the type of photograph used, human experts assigned "point 7" (granting disability certification benefits) slightly more frequently than the language models ($U = 539$, $p = .043$), although this difference was small ($r = .23$). The values assigned in the assessments of the level of functioning were otherwise comparable. Neither among the experts nor among the language models was a statistically significant relationship found between the assessment of the level of functioning and the number of assigned "point 7"; in both groups, these variables were unrelated. This indicates a lack of consistency in judgments, as it might be expected that the rate at which long-term care is granted would rise and fall in line with a decrease or increase in the assessed level of functioning (Fig. 5).

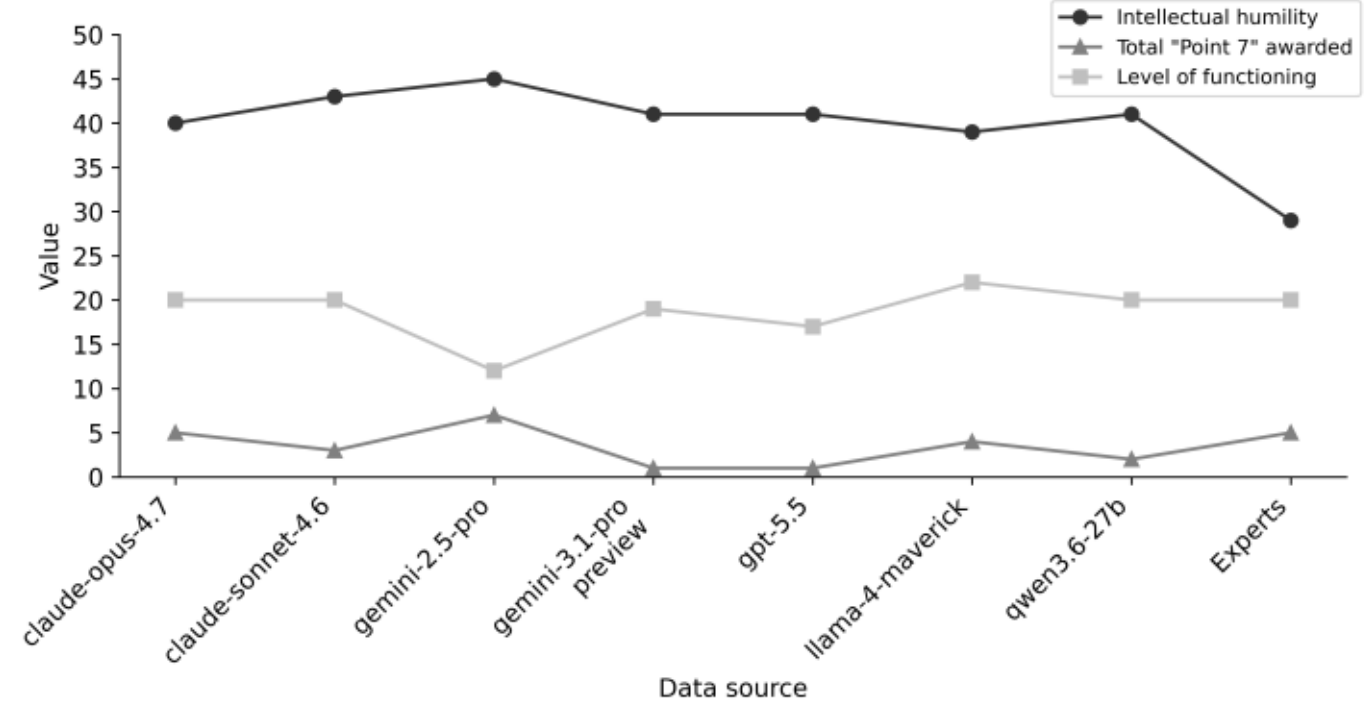

**Figure 5:** Overall Comparison of Assigned "point 7", Level of Functioning Ratings and Intellectual Humility for Experts and Language Models.

On the measure of intellectual humility, the language models achieved substantially higher and more homogeneous scores ($M = 41.43$, $SD = 1.99$) than the experts ($M = 29.03$, $SD = 8.05$) (Fig. 5 $U = 241$, $p < .001$, $r = .62$). Nevertheless, in neither group was the level of intellectual humility associated with the type of decision made, whether measured by the number of assigned "point 7" or by the assessment of the level of functioning. In practical terms, the higher intellectual humility reported by LLMs did not translate into more coherent judgments. Models that expressed greater willingness to acknowledge uncertainty, recognize that their conclusions might be mistaken, and reconsider their beliefs did not show a stronger correspondence between their assessments of functional ability and their decisions regarding the need for long-term assistance. The same pattern was observed among human experts: individuals with higher self-reported intellectual humility did not make decisions that were more closely aligned with their functional assessments. Thus, declarative intellectual humility, as measured by the SIHS, was not associated with decision consistency in either group. This finding suggests that expressing epistemic caution or acknowledging the possibility of error does not necessarily translate into more consistent decision-making when applying complex, value-laden criteria in practice.

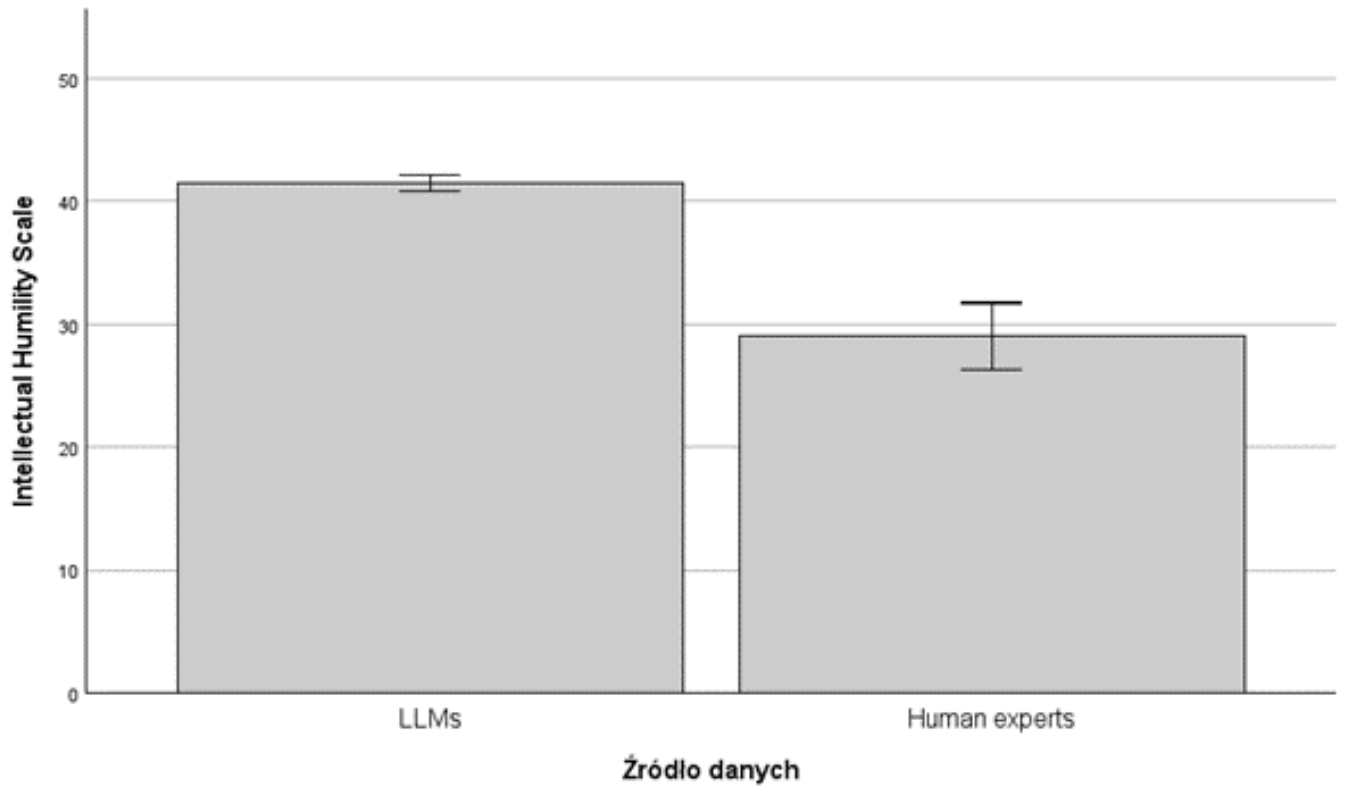

**Figure 6:** Mean Intellectual Humility Scores (M ± SD) for LLMs and Human Experts.

Similarities in decision-making patterns Support decisions differed among LLMs, physicians, and psychologists ($H(2) = 8.92$, $p = .012$, $\eta 2 = .09$), with descriptive statistics for each group presented in Table 1. Post hoc comparisons indicated that LLMs did not differ significantly from physicians in the frequency of granting support; however, they granted support significantly less frequently than psychologists ($U = 180.50$, $p = .003$, $r = .34$). Thus, although LLMs demonstrated a decision pattern comparable to that of physicians, their judgments differed from those of psychologists, who were more likely to classify individuals as requiring support. Within the expert group, physicians and psychologists assessed the level of functioning of all individuals described in the case materials similarly, indicating that both professional groups reached comparable evaluations of functional ability. However, despite similar assessments of functioning, physicians assigned “point 7” significantly less frequently than psychologists ($t(31) = 2.19$, $p = .036$, $d = 0.79$). This difference suggests that the two professional groups had different understandings of "basic life needs”. Source N “Point 7” M (SD) LLMs 21 3.57 (2.68) Physicians 18 3.67 (2.00) Psychologists 17 5.35 (2.21).

| **Source** | **N** | **“Point 7” M (SD)** |
|---|---|---|
| LLMs | 21 | 3.57 (2.68) |
| Physicians | 18 | 3.67 (2.00) |
| Psychologists | 17 | 5.35 (2.21) |

**Table 1:** The number of Assignments of Disability Certification Benefits ("Point 7") across LLMs, physicians, and psychologists.

### Qualitative insight into the key concepts

During the interviews, the experts explained their understanding of the basic concepts and the key factors taken into account when making decisions. For the models, the post-hoc self-rationalisation technique was used (Madsen, Chandar, and Reddy 2024; Marasovic et al. 2022). To further contextualize these qualitative findings, we analysed transcripts of semi-structured interviews conducted with all expert participants and compared them to the models’ answers. In these interviews, most psychologists defined “basic life needs” broadly, referring not only to physiological requirements but also to psychological and social dimensions, including emotional, relational, and communication needs. By contrast, most physicians defined “basic life needs” mainly through reference to physiological functioning and independent performance of essential self-care activities, such as getting out of bed, walking, preparing meals, eating, and dressing. These differences in conceptualization may explain why psychologists more frequently assigned “point 7” despite reaching similar evaluations of overall functioning. The psychological and social needs considered by psychologists were included in the definitions provided by only two models and exclusively in the context of expressing basic needs and maintaining physical safety (i.e., avoiding threats to life and health). Although one model – Claude Sonnet 4.6 – included “communication needs,” defining them as “the ability to express one’s own needs, understand others, and communicate with one’s environment,” this was not reflected in the number of “yes” responses it provided.

## Conclusions

Despite the fact that neither experts nor LLMs showed significant susceptibility to the anchoring and representativeness manipulations, our findings suggest that the more consequential challenges lie in the decision inconsistency (the lack of coherence between functional assessments and supporteligibility decisions) and epistemic reductionism (the deficitladen interpretive framework reproduced by language models). Despite comparable assesment of the level of functioning, physicians were less likely than psychologists to indicate that an individual required constant or long-term care or assistance from another person due to a significantly limited ability to live independently, while LLM decisions resembled those of physicians. This pattern was reflected in how the groups operationalized the value-laden concepts underlying their judgments. Most LLMs, like physicians, interpreted “the ability to live independently” and “basic life needs” primarily in terms of biological requirements and self-care, including eating, personal hygiene, dressing, mobility, and other activities of daily living. This suggests that AI used for supporting disability decisions risks collapsing evaluations into physical activity deficits, systematically denying support to patients whose restrictions concern communication and interpersonal participation (which are crucial in the WHO ICF framework, (Anner et al. 2012)). Psychological, communicative, and social needs emphasized by psychologists appeared in only two models’ definitions and mainly in relation to communicating basic needs or maintaining physical safety. Claude Sonnet 4.6 additionally recognized communication as the ability to express needs, understand others, and interact with one’s environment, but this broader definition did not translate into more frequent support decisions. Because opinions diverging from physicians’ assessments may receive less weight in disability-assessment practice, LLMs could potentially provide an independent second opinion (Erden, Hummerstone, and Rainey 2021; Lai et al. 2025). Our results indicate, however, that current models may instead reinforce the dominant medical framing rather than broaden the epistemic basis of assessment (Lutte et al. 2024). Improving resistance to familiar cognitive heuristics may therefore be insufficient to ensure fairer decisions (Croskerry and Norman 2008). Psychiatric diagnosis also carries significant political weight, as it shapes not only how mental distress is understood but also how individuals are socially categorized, governed, and granted or denied access to rights, resources, and institutional support (Bizzari and Brencio 2024). Audits of high-stakes

AI should also examine decision coherence and the normative frameworks encoded by models. The observed reductionism may originate in dominant patterns within pretraining data or be reinforced through post-training alignment. Future research could distinguish these sources by comparing base and instruction-tuned models, examining successive training checkpoints, and conducting controlled fine-tuning on medical, biopsychosocial, and neurodiversity-affirming corpora, which would also take into account the first-person perspective on the experiences of neurodivergent people (Dudzic et al. 2026).

## Limitations

This research has several limitations. The first is that measuring language models intellectual humility with self-report questionnaries designed for humans may result in artifacts with no relationship to actual bias resistance. The second is that it is not known whether human experts and language models showed no susceptibility to anchoring and representativeness heuritics due to genuine lack of susceptibility or because the visual and textual manipulation was too subtle. The third limitation is the small size of both samples, which reduced the experiment's statistical power. The fourth, regarding language models, is that the study relies on 3 interations at default temperature and without in-context learning, such as adapting biopsychosocial perspective. It is also possible that persona-prompting would shift the models' assessment of basic life needs away from the medical framing.

## Ethical Considerations

All case vignettes used in the study were entirely fictional. The human-participant component, which involved deception, received approval from the Research Ethics Committee of the first author's home institution (project no. anonymized). All participants provided informed consent before taking part in the study.